\documentstyle{mn}
\title[JVAS/CLASS polarization - I. Data and catalogue production]
{A survey of polarization in the JVAS/CLASS flat-spectrum radio source
surveys: I. The data and catalogue production \\}
\author[N. Jackson, R.A. Battye, I.W.A. Browne, S. Joshi, T.W.B. Muxlow, P.N. Wilkinson]
{N. Jackson, R.A. Battye, I.W.A. Browne, S. Joshi,\newauthor T.W.B. Muxlow, P.N. Wilkinson\\
University of Manchester, Jodrell~Bank~Observatory, Macclesfield, Cheshire, SK11 9DL U.K.\\}

\begin{document}
\input{psfig}
\label{firstpage}
\maketitle

\begin{abstract}

We have used the very large JVAS/CLASS 8.4-GHz surveys of flat-spectrum radio
sources to obtain a large, uniformly observed and calibrated, sample of
radio source polarizations. These are useful for many investigations of
the properties of radio sources and the interstellar medium. We discuss
comparisons with polarization measurements from this survey and from
other large-scale surveys of polarization in flat-spectrum sources.

\end{abstract}

\begin{keywords}
techniques:polarimetric -- radio continuum:general -- surveys -- galaxies:active
\end{keywords}

\section{Introduction}

Polarization measurements are potentially important because they
provide information about magnetic fields and the media between source
and observer. Radio emission from active extragalactic objects is
generated by relativistic electrons spiralling around magnetic field
lines, either close to the active centre or in the radio jets ejected
from it.  Radio waves passing through a magnetised plasma suffer
Faraday rotation whose magnitude depends on the magnetic field and
electron density along the path; this can occur within the
synchrotron-producing region, in the interstellar medium of the
emitting galaxy or in passage through our own Galaxy. It is also
important to measure the polarization of large numbers of discrete
radio sources in order to remove their contribution from future
measurements of the polarization of the cosmic microwave background radiation.

Studies of radio polarization in extragalactic sources have yielded
important results in the past, including support for orientation
models of active galaxies (Laing 1988; Garrington et al. 1988). They
have also been used to investigate the physics of radio jets on
kiloparsec scales, which can be polarized by tens of percent and whose
polarization structure can give information on magnetic fields and jet
confinement (e.g. Laing et al. 2006; Leismann et al.  2005; Hardee 2003;
Gizani \& Leahy 2003; Ferrari 1998; Laing 1996) down to parsec scales 
close to the central black
hole (e.g. Lister \& Homan 2005; Hughes 2005; Lyutikov, Pariev \& Gabuzda
2005; Fraix-Burnet 2002).

The subject of this paper is the radio polarization of core-dominated,
flat-spectrum, radio sources. At flux density levels of $>$100~mJy at
frequencies of a few GHz, these sources are predominantly quasars and
BL Lac objects in which the bipolar jets are oriented close to the
line of sight. Consequently, the radio emission from the approaching
jets is enhanced by Doppler beaming, leading to associated observable
effects such as strong variability and apparent superluminal motion
caused by the relativistic motion close to the line of
sight. The integrated polarization in these flat-spectrum sources is typically a few
per cent (e.g. Saikia, Kodali \& Swarup 1985; Okudaira et
al. 1993). The physics is somewhat complicated, since flat-spectrum
``cores'', when observed with milliarcsecond resolution, consist of
multiple synchrotron components of different sizes and hence with
different frequencies at which the emission becomes optically thick
and below which the spectrum suffers a self-absorption turnover. The
integrated rotation measure is usually quite low, often less than a few
tens of rad~m$^{-2}$ (e.g. Rudnick \& Jones 1983, but see also Zavala \&
Taylor 2004).

Over and above these astrophysical inferences, occasional attempts
have been made to investigate the hypothesis that observations of
apparently ordered polarizations can be used to make cosmological
inferences. Amongst these are chiral effects on the propagation of
light (Nodland \& Ralston 1997; see also Carroll \& Field 1997),
coupling of light pseudoscalars, such as axions, to the photons
(Raffelt \& Stodolsky 1988), and universal rotation (Birch 1982). More
recently, Hutsem\'ekers et al. (2005) have claimed alignments on large
angular scales of the optical polarization position angles of quasars.
Our data can be used to place constraints on effects of this nature
and these are discussed in more detail in Paper II of this series.

In this paper we describe the data calibration and analysis for the
largest survey of polarization in compact radio sources to date,
namely the JVAS and CLASS surveys (Patnaik et al. 1992; Browne et
al. 1998; Wilkinson et al. 1998; Myers et al. 2003; Browne et
al. 2003). We discuss the steps that need to be taken to ensure clean
polarization measurements with low systematics in these large
samples. In subsequent papers we address uses of this sample. In Paper
II we discuss cosmological alignments prompted by the claims of
Hutsem\'ekers et al. (2005) of extreme-scale alignments of quasar
optical polarization vectors. In Paper III we will present further
polarization observations of a subset of the JVAS/CLASS sources.

\section{Observations and data analysis}

\subsection{The JVAS/CLASS surveys}

The prime motivation for JVAS (the Jodrell-VLA Astrometric Survey) was
to identify sources suitable for use as interferometer phase
calibrators and to measure accurate positions for them. A secondary
aim was to look for any sources that might have been gravitationally
lensed. For both these objectives, pre-selecting sources which one
might expect {\it a priori} to have compact radio structures was
desirable. For this reason only those sources with known flat
radio-spectra ($S^{\rm  5~\tiny GHz}_{\rm  1.4~\tiny GHz}>-0.5$) 
were selected. Spectral selection for JVAS was initially carried out 
using the 5-GHz 87GB survey (Gregory \& Condon 1991) and the 
1.4-GHz Green Bank survey (White \& Becker 1992).
Each JVAS source was then observed for 2~min
with the VLA in its A-configuration at a frequency of 8.4~GHz. A full
description of the sample selection process and the observational
details can be found in Patnaik et al. (1992). In three separate 
observing sessions, each
separated by around 15 months, 2720 sources stronger than
$\sim$200~mJy at 5~GHz were successfully observed in the region of sky with
declination $\geq 0^{\circ}$ and $\mid$b$\mid \geq 2.5^{\circ}$. The
observations in each session were typically divided into blocks of time
lasting for $\geq$12~hr, each of which was calibrated and analysed
separately.

The CLASS survey (Cosmic Lens All-Sky Survey) consisted of 30-second
snapshot observations at 8.4~GHz of all sources with a 5-GHz flux
density of $>$30mJy which had not already been observed as part of
JVAS. In addition, some sources below 30~mJy were observed, down to
the GB6 survey limit of $\sim$20~mJy and using newer 
surveys (GB6, Gregory et al. 1996; WENSS, Rengelink et al. 1997; NVSS,
Condon et al. 1998). Full observing details and a more complete
description of sample selection can be found in Myers et
al. (2003). Four observing sessions in 1994, 1995, 1998 and 1999
covered the whole sky between declinations of 0$^{\circ}$ and
70$^{\circ}$, with the more southerly regions being observed later in
the programme. 13783 sources were observed, making a total of 16503
sources for the whole JVAS/CLASS programme.

\subsection{Calibration}

The original purpose of the CLASS survey was to identify gravitational
lens systems. The analysis was done using a {\sc difmap} script
(Shepherd 1997) for automatic mapping and identification of
candidates.  For the present work we have re-edited, recalibrated and
re-imaged the datasets as described below. This is because we have
found that bad data on individual telescopes or calibration errors
which do not cause major problems for lens identification can become
serious impediments to accurate polarization calibration and
mapping. All analysis was done using the NRAO Astronomical Image
Processing System ({\sc aips}).  Because we want to be able to perform
statistical analysis on large samples of sources with low levels of
polarization, the polarization calibration is crucial, and we
therefore describe it in some detail. Errors in the antenna
polarization calibration at one epoch of observations will show up as
correlated polarizations with similar position angles in the patch of
sky observed at that epoch. 

After careful editing of bad data, phase and amplitude solutions were obtained
using point-source phase calibrators which were typically observed
every 15-20 minutes during the CLASS observations, and more frequently 
in the case of JVAS. These solutions were inspected,
edited where necessary to remove discrepant points, and interpolated
before application to the data on the target sources. 

\subsubsection{Instrumental polarization terms}

Instrumental polarization terms were then estimated using the {\sc
aips} task {\sc pcal}. For this process it is necessary to find a
calibrator which either has a known polarization, or has an unknown
polarization but which has been observed at a range of parallactic
angles; this allows its polarization at the time of observation to be
calculated. For the majority of epochs of CLASS data the unpolarized 
source 3C84 (J0319+415) was observed. Aller, Aller \& Hughes (2003) 
present monitoring results for 3C84 over 17 years and quote 
a 14.5-GHz polarization of 0.12\%$\pm$0.01\%, which is well within 
the error to which we are able to determine instrumental terms. 
We have therefore assumed 3C84 to be completely unpolarized. In some 
cases we have used OQ208 as a zero-polarization calibrator which has 
been observed on numerous occasions with the WSRT (A.G. de Bruyn, 
private communication). In a few cases, particularly in many of the 
CLASS observations from 1994, we do not have a suitable instrumental
term calibrator and have therefore excluded these observations
from our analysis. This results in a sparse sampling of some of the
sky, in particular for areas north of declination 45$^{\circ}$.

For one observing session in 1991, we have observations of both the
unpolarized calibrator OQ208 and observations of a strong 
polarized calibrator, B1611+343, taken at a range of parallactic angles. In
Figure 1 we show the instrumental terms derived from each of these two
observations. They are entirely consistent with each other giving us
confidence in the use of OQ208 as an unpolarized calibrator. Figure 2
shows the resulting polarizations derived for 150 target sources using
the two methods. Little difference is evident. If significantly
polarized sources are considered (polarized flux density $>$1mJy), the
rms deviation in position angles between the two calibrations is 
6$^{\circ}$. We have also experimented with the use of the solution 
interval for the determination of instrumental term solutions (between
15s and 60s, a standard value of 30s being eventually chosen), 
as well as choice of
reference antenna, and find that the result is robust to these
choices. In practice, the mean instrumental term for each
telescope was normally between 1\% and 2\%. Excursions outside this
range could nearly always be traced back to bad data in one or more
telescopes which upsets the overall solution.

\begin{figure*}
\begin{tabular}{cc}
\psfig{figure=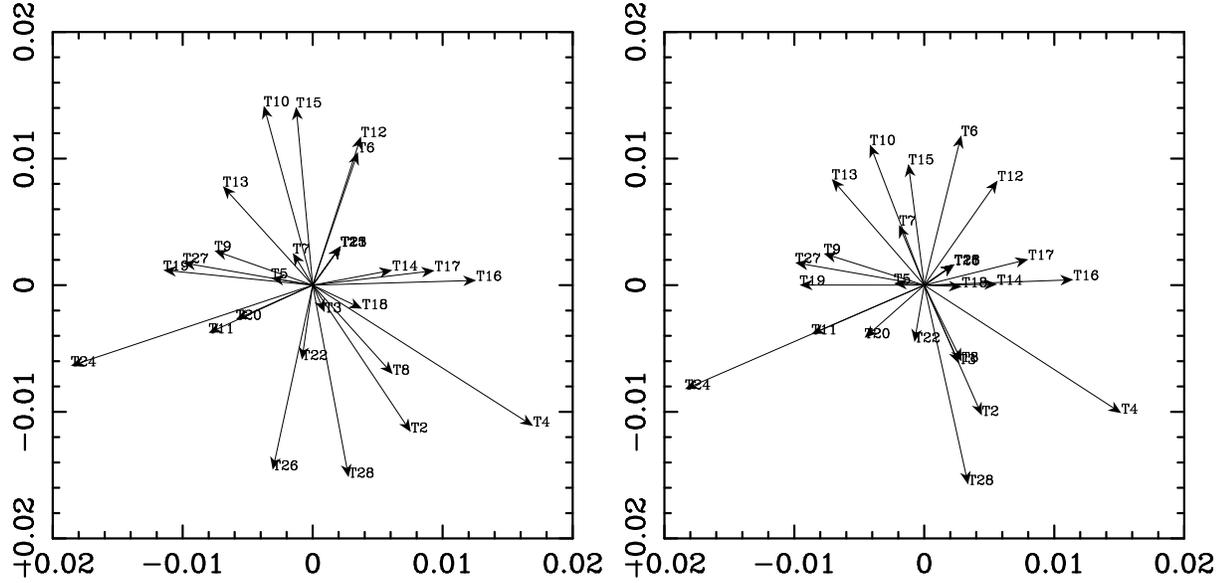,width=16cm,angle=-90}
\end{tabular}
\caption{\small Instrumental polarization terms per telescope for the
observations taken in 1991 as part of the JVAS survey. The scales on
the axes run from $-3$\% to 3\% of instrumental polarization. The length
of each line gives the amplitude of the instrumental term; that is, the amplitude
of polarization that this telescope would record for an unpolarized
source. The angle gives the phase in such a way that a complete rotation
on the diagram would rotate the polarization vector by 180$^{\circ}$.
The number attached to each arrow refers to the specific VLA antenna. On the
left is the calibration achieved by using the polarized source 1611+343
observed at different parallactic angles, and on the right is the
calibration made using the assumption that the source OQ208 is
unpolarized. Note the similarity in the derived instrumental terms.}
\end{figure*}

\begin{figure*}
\psfig{figure=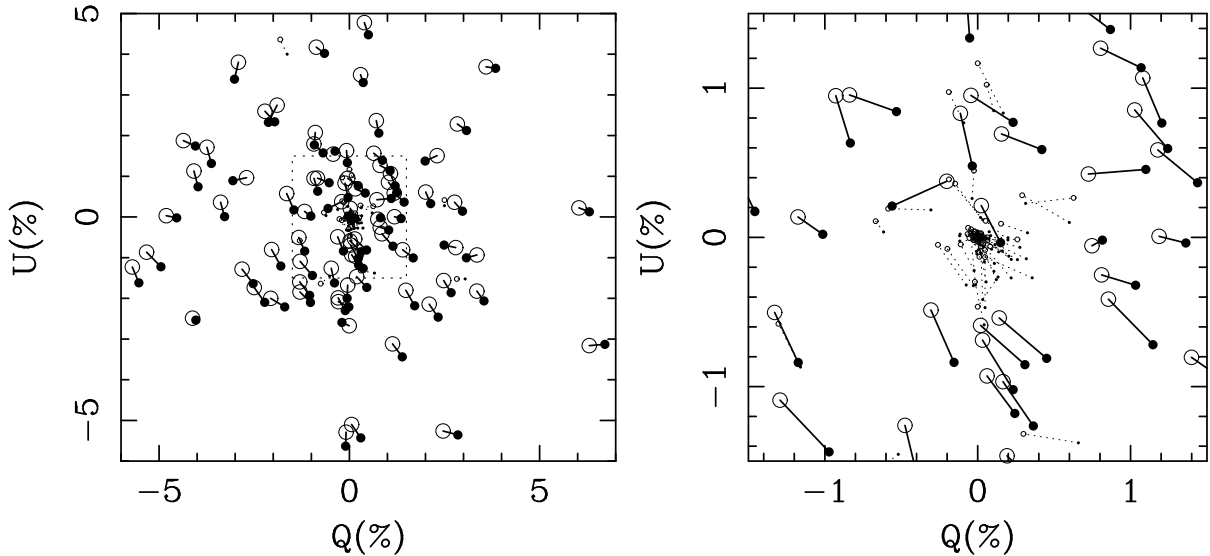,width=16cm,angle=-90}
\caption{\small Comparison of derived polarization of 150 
sources obtained using two different
calibrations of instrumental polarization in observations from
1991 June 16. Axes represent $Q$ and $U$ as a percentage of Stokes $I$. 
Unfilled and filled circles represent calibrations using the
polarized source 1611+343 and the unpolarized source OQ208. Larger
symbols represent significant detections of polarization ($>$1~mJy in
polarized flux density). The right-hand panel is a blowup of the inner
region of the left-hand panel, indicated by a dotted line in the
left-hand panel.}
\end{figure*}

\subsubsection{Polarization angle determination}

After the instrumental terms were determined, the corrections were applied to 
observations of a polarized source of known position angle in order 
to determine the relative phase of the left-hand and right-hand 
polarization channels. This is necessary in order to rotate the 
phases for all baselines within each polarization channel so that the
average phases on the left-hand and right-hand channels are
consistent with each other and also with the known
linear polarization angle of the position-angle calibrator. We used
3C286 (position angle 33$^{\circ}$) and 3C48 (position angle
$-70^{\circ}$) for this purpose (Myers \& Taylor 2006).

We have checked that the phase rotations applied to left- and right-hand
data in each epoch within a season do not differ significantly, although
some drift is expected resulting from system changes and which the phase
rotation is designed to remove. 
In most cases this was true to within 20$^{\circ}$ (r.m.s. deviation
between epochs was 19$^{\circ}$ in 1995 and 14$^{\circ}$ in 1998),
apart from a significant change which occurred between 1998 April 5
and 1998 May 21. 

When the instrumental polarization term correction and the L-R phase rotation are
both determined, a polarization angle can be derived for each baseline from
observations of the polarized calibrator. Ideally, there should be a
small scatter between baselines of less than 10$^{\circ}$ in these 
measurements, giving a very small error in the overall polarization
angle averaged over $~\sim$300 baselines. In some epochs, all from the 
CLASS part of the survey, this desirable level of scatter is exceeded
by factors of 3-4. However, in such epochs where two polarized
calibrators (3C48 and 3C286) are available, we can determine the L-R
phase corrections from one source and use them to attempt to recover the
polarization angle for the other source. In these cases, and in other
cases where two polarized calibrators are available, the angle is
recovered to within 10$^{\circ}$ of that expected, suggesting that there is
no major systematic problem in the use of data from these epochs and in
particular that derived polarization angles have systematic calibration
errors of $\leq 10^{\circ}$.
In most of the survey, including all the JVAS observations and the CLASS
observations from 1998 March and April, the reliability is likely to be
significantly better than this.

Table 1 shows the JVAS/CLASS epochs with the calibrators used for each,
together with the important polarization parameters which have been
derived for each epoch.

\begin{table*}
\begin{tabular}{lccccccp{45mm}}
Epoch & \multicolumn{2}{c}{Sky area} & Zero-pol& Instrumental &
Angle & L$-$R residual/$^{\circ}$ & Comments\\
&RA&Dec&calibrator&pol. term/\%&calibrator&&\\
&&&&&&&\\
900219 & all & $>$35 & 3C84 & 1.0\% & 3C48 & 2 & \\
910616 & all & 20--35 & OQ208 & 1.0\% & 3C286 & 5 & \\ 
921017 & all & $<$20 & 3C84 & 1.1\% & 3C286 & 2 & \\
940301 & 10-20h & 45-53 & 3C84 & 1.6\% & 3C286 & 3 & \\
940404 & 7-10h & 45-70 & 3C84 & &&&\\
      & \&14-20h & &&&&&\\
950813 & 6-15h & 29-36 & 3C84 & 1.0\% & 3C286 & 16  &Calibrator from 950814 used\\
       & 6-9h & 45-70 &&&&&\\
950814 & 0-5h & $>$37 & 3C84 & 1.5\% & 3C286 & 24 & \\
       & 6-10h & 37-45 &&&&&\\
950829 & 20-6h & $<$10 & 3C84 & 1.3\% & 3C48 & 43 & \\
950902A & 20-6h & 10-17 & 3C84 & 1.3\% & 3C48 & 16 & \\
        & 6-20h & various&&&&&\\
950902B & 20-6h & 17-30 & 3C84 & 1.3\% & 3C48 & 31 & 3C286 agrees to
$<$10\% in R-L offset despite large scatter\\
980314A & 2-24h & 8-17 & 3C84 & 1.3\% & 3C286 & 2 & \\
        & 2-12h & $<$8 &&&&&\\
980314B & 1-11h &17-28 & 3C84 & 1.0\% & 3C286 & 2 &Calibrator from 980314A used\\
980403 & 12-24h & $<$7 & 3C84 & 1.3\% & 3C286 & 4 & \\
       & 10-19h & 17-30 &&&&&\\
       & 17-13h & 28-45 &&&&&\\
980510 & 20-24h & 17-28 & OQ208 & 1.5\% & 3C48 & 44 &Calibrator from 980521 used\\
       & 0-2h & $<$8&&&&&\\
980521 & 13-17h & 28-45 & OQ208 & 1.2\% & 3C286 & 14 & \\
       & 8-21h & 45-70 &&&&&\\
990816 & 12-20h & 0-20 & 3C84 & 1.3\% & 3C286 & 45 & \\
       & various & various &&&&&\\
\end{tabular}
\caption{Observing epochs for JVAS/CLASS. For each epoch we list the
approximate sky area principally covered by the observation. The
zero-polarization calibrator observed to calibrate the instrumental
polarization terms is listed together with the derived instrumental polarization
in columns 4 and 5; in most cases these can be brought
below 1.5\% with careful editing of bad data. The amplitudes of these
terms are also given by the lengths of the vectors in Fig. 1. 
Column 6 lists the
polarized calibrator used to determine L-R phase offsets as described in
the text. The scatter between measurements of polarization angle for
different baselines, derived from applying the instrumental polarizations
to the polarized calibrator, varies widely between epochs.}
\end{table*}

\subsection{Imaging and the derivation of position angles}

The calibrated datasets were mapped using an automatic procedure
implemented using an {\sc aips} runfile. A shift was applied to the
phases of the $u-v$ data in order to centre the observations on the
brightest point source found in the earlier automated analysis
described in Myers et al. (2003). Sources were not included if
they were not detected in the previous analysis, 
or if a source was
located but judged to be extended on the basis of angular sizes
(defined as being fitted by a Gaussian FWHM  of 0\farcs5 or greater). 
An initial map of 12\farcs8$\times$12\farcs8 was produced using natural
weighting. If the source had previously been detected by the Myers et
al. (2003) analysis with a flux density of 40mJy or greater, the initial
map was used as an input for phase self-calibration using clean
components brighter than the first negative;  in cases where the
flux density is greater than 80mJy one iteration of amplitude
self-calibration was also applied, using the results from the initial
phase self-calibration as an input model. For the purposes of visual
inspection, final images in Stokes $I$, $Q$ and $U$ were then made using the
{\sc clean} algorithm (H\"ogbom 1974) 
using the task {\sc aips imagr}, and combined
into polarization level and position angle maps.

Polarization parameters were extracted using fits to the processed
(and in the case of strong sources, self-calibrated) $u-v$ Stokes $Q$
and $U$ data using the task {\sc aips uvfit}. In order for this
algorithm to converge successfully, the position of the centre of
polarized flux is required. This was calculated by making a
total-intensity map and using the brightest point in this map as the
$Q$ or $U$ centre, on the assumption that these coincide.

We did not use the value derived from applying the CLEAN algorithm to
the $Q$ and $U$ data, as this can cause the measured flux densities to
be decreased by CLEAN bias (e.g. Condon et al. 1998). The
model-fitting procedure is valid provided that there is no extended
polarized structure such as a polarized radio jet, in which case it
provides an averaged value. 


We checked for the overall uniformity of the distribution of the
model-fitted polarization position angles and found, as expected, that
the position angles derived from direct fits to the data are
statistically uniform in the range from 0$^{\circ}$ to
180$^{\circ}$. In contrast, angles derived from CLEANed $Q$ and $U$
data show peaks at 0$^{\circ}$, 45$^{\circ}$, 90$^{\circ}$ and
135$^{\circ}$, corresponding to CLEAN bias forcing the $Q$ and $U$
fluxes towards zero. This effect of CLEAN bias on polarization
determinations is unexpected and potentially important; it will be
discussed further in a future paper (Battye et al., in preparation).

In order to keep only significant polarization detections, we imposed a
limit of 1~mJy in polarized flux in most subsequent analysis, 
corresponding to approximately 4$\sigma$; below these levels the derived
polarized flux is increasingly affected by positive bias (e.g. Simmons \&
Stewart 1985). However, the polarization parameters for all sources are 
available in the electronic version of this paper. An extract from the 
first page of the data table is shown in Table 2 and includes the position
of the source, the epoch of the data and the derived flux densities in 
Stokes $I$, $Q$ and $U$ together with the polarization percentage and 
position angle. The total number of sources in the catalogue with 
polarization measurements is 12555 of the total JVAS/CLASS sample of 16503.
Of these, 4294 have significant detections of polarization, with a polarized
flux density $\geq$1~mJy.

\begin{table*}

\begin{verbatim}

--------------------------------------------------------------------------------
RA (hh mm ss)  Dec (dd mm ss)      I    SigI     Q    SigQ     U    SigU  PA/deg
--------------------------------------------------------------------------------
00 00 07.0341  +08 16 45.040      29.4   0.2   -0.26  0.19   +0.28  0.19   66.2
00 00 10.0908  +30 55 59.420      20.0   0.3   +0.18  0.20   -1.40  0.20  138.7
00 00 14.8771  +27 51 57.577      22.8   0.3   +0.35  0.21   +0.11  0.21    9.1
00 00 19.2833  +02 48 14.657      81.8   0.3   +1.21  0.32   -1.66  0.33  153.1
00 00 19.5679  +11 39 20.718      28.9   0.3   +0.40  0.27   +0.32  0.26   19.4
00 00 26.8395  +44 31 11.700       5.4   0.2   -0.27  0.20   -0.14  0.21  103.8
00 00 27.0230  +03 07 15.635      96.0   0.4   +1.87  0.37   -0.37  0.36  174.5
00 00 30.1085  +47 16 43.313      16.3   0.3   -0.03  0.24   +0.09  0.24   53.1
00 00 35.1294  +29 14 35.823      64.5   0.2   +0.48  0.25   -0.30  0.25  164.1
00 00 44.3279  +03 07 54.199      60.6   0.3   +0.86  0.32   -0.36  0.32  168.6
00 00 49.7361  +32 52 57.109      13.5   0.3   -0.39  0.28   +0.46  0.28   65.1
00 00 56.0910  +25 16 20.152      25.2   0.4   -0.96  0.35   +0.13  0.35   86.2
00 01 07.8693  +24 20 11.799      51.4   0.3   -3.18  0.28   +0.66  0.29   84.1
00 01 09.5365  +02 43 09.588      60.3   0.2   -0.30  0.18   +0.50  0.18   60.5
00 01 14.3441  +06 14 22.011      20.8   0.3   -0.02  0.33   +0.35  0.34   46.4
00 01 14.8643  +23 58 10.617     117.1   0.5   -0.30  0.36   -0.25  0.36  110.1
00 01 21.6723  +25 26 55.519      41.2   0.3   -0.78  0.29   -0.40  0.29  103.5
00 01 23.6973  +06 32 30.966      33.9   0.3   +0.88  0.19   -0.52  0.19  164.8
00 01 32.2272  +13 52 58.482      13.1   0.3   +0.20  0.25   -0.52  0.26  145.6
00 01 32.3700  +21 13 36.216      99.2   0.4   -0.83  0.32   +1.16  0.32   62.8
00 01 34.4532  +07 23 12.903      71.0   0.3   +0.86  0.35   +2.63  0.35   36.0
00 01 43.4710  +07 01 23.570      50.3   0.3   +0.11  0.22   +0.38  0.22   37.1
00 01 46.2388  +46 16 32.368      15.7   0.2   -0.18  0.21   -0.32  0.21  120.8
--------------------------------------------------------------------------------
\end{verbatim}

\caption{\small The first page of the polarization angle measurements
with the automated analysis as described in the text. The first six 
columns are the position in RA (3 columns) and declination (3 columns)
of the major component as derived from the CLASS analysis (Myers et al. 2003).
Remaining columns contain measurements derived from the direct fits to the
$u-v$ data. Flux densities in milliJanskys follow for total flux (Stokes $I$), 
error in $I$, Stokes $Q$, error in $Q$, Stokes $U$ and error in $U$. 
The final column is the derived position angle in degrees 
($\frac{1}{2}\tan^{-1}(U/Q)$).}
\end{table*}

\subsection{External comparisons with other data.}

Do the CLASS\footnote{In this and subsequent sections we refer to the
combined JVAS/CLASS surveys simply as ``CLASS''.} 
polarization position angles agree with other
measurements? We have made comparisons, some direct and some indirect,
that convince us that our CLASS polarization angles are reliable.
Okudaira et al. (1993) have used the Nobeyama 45-m telescope to
measure the polarizations of 99 flat spectrum radio sources at 8.4~GHz
and their sample has 76 objects in common with ours. We plot the
histogram of position-angle differences in Fig. 3. There is a clear
peak around zero degrees, with a rms dispersion of
$\pm40^{\circ}$. Some spread is expected because many of the sources
are known to be extremely variable both in their total intensity and
polarization properties, position angles (e.g. BL Lac, 3C345, 3C454.3,
etc.).  In addition, the angular resolutions of the two sets of
observations are very different. Resolution may make a difference
because emission from optically thin, kiloparsec-scale jets, is
generally more highly polarized, and sometimes at a different position
angle, compared to that from the compact cores (e.g. Saikia \& Salter
1988).

Another test is to compare the CLASS position angles with the intrinsic 
position angles, corrected for Faraday rotation, determined by Broten et
al. (1988). Typical rotation measures arising from a combination of Galactic and
internal Faraday rotation are $\sim$20~rad~m$^{-2}$ which means that
most of the CLASS measured angles should be close to the intrinsic
ones. The difference between the CLASS angles and those from Broten et 
al. (1998) for the 67 objects in common are also shown in Fig. 3. The 
spread is about the same ($\sim 40^{\circ}$) as that of the
Okudaira et al. (1993) sample but there is a peak which a formal average
of the data shows to be centred around $(-7\pm 5)^{\circ}$.

A typical rotation measure of $\sim$ 20 rad m$^{-2}$ will produce a
rotation of $\sim 50^{\circ}$ at a wavelength of 21~cm which,
if left uncorrected, would effectively smear out any record of the
intrinsic position angle. There are, however, areas of sky at high
Galactic latitudes where average rotation measures are significantly
less than 20 rad m$^{-2}$ and hence one would expect to be able to see
a correlation between the CLASS position angles and those measured at
21~cm in these regions. The NVSS survey, which was made at a wavelength 
of 21~cm, lists polarization information on all the sources (Condon et
al. 1998). In Fig. 3 we show the histogram of position angle
differences between CLASS and NVSS for the 1424 common sources in the
region $\mid$b$\mid \geq 30^{\circ}$ and which have NVSS polarized
flux density $\geq$2~mJy and CLASS polarized flux density
$\geq$1~mJy. As expected there is a significant peak near zero
degrees, although the mean value of the data is actually located at
(3.9$\pm$1.2)$^{\circ}$.
   
\begin{figure}
\begin{tabular}{c}
\psfig{figure=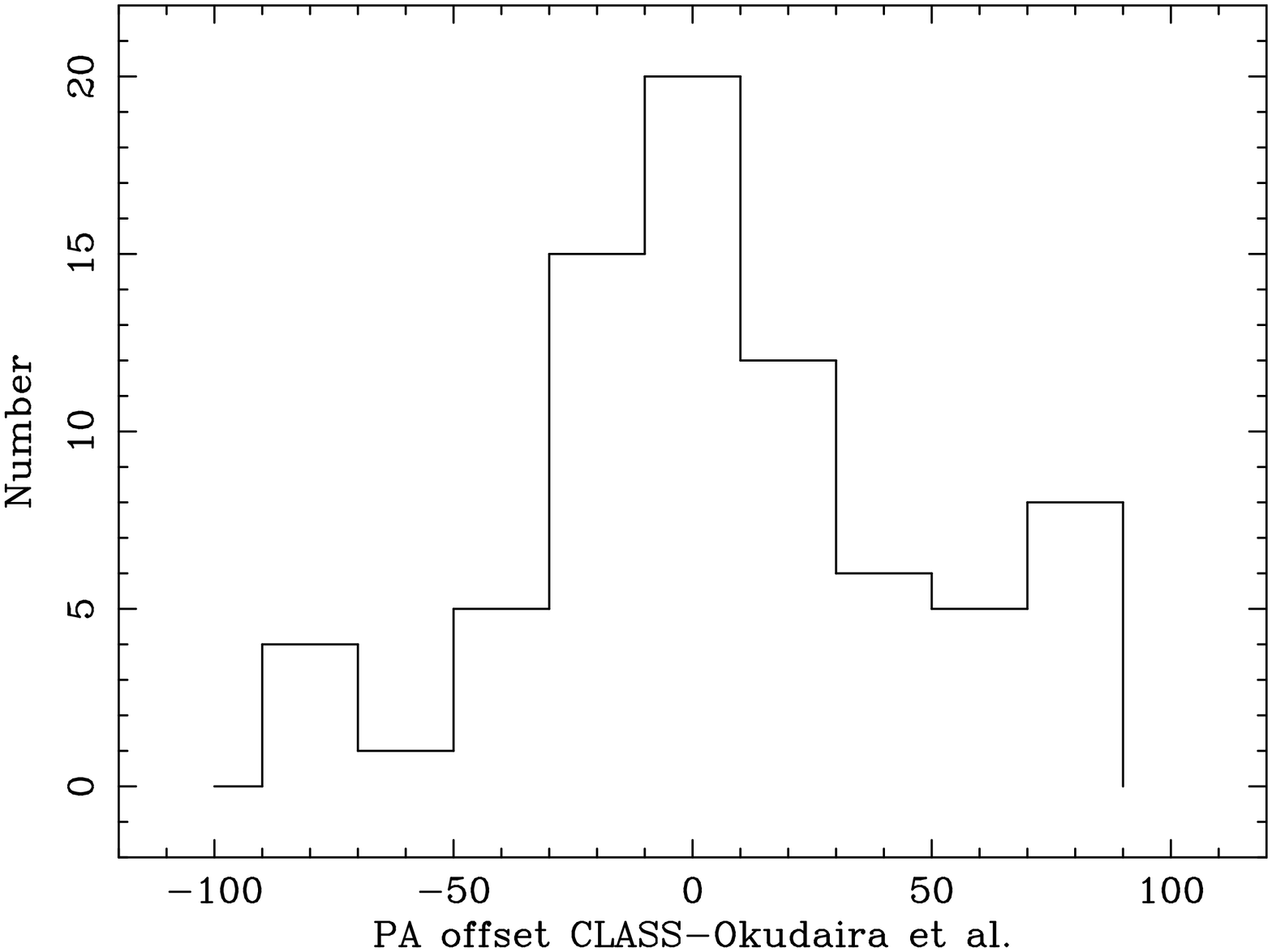,width=8cm,angle=-90}\\
\psfig{figure=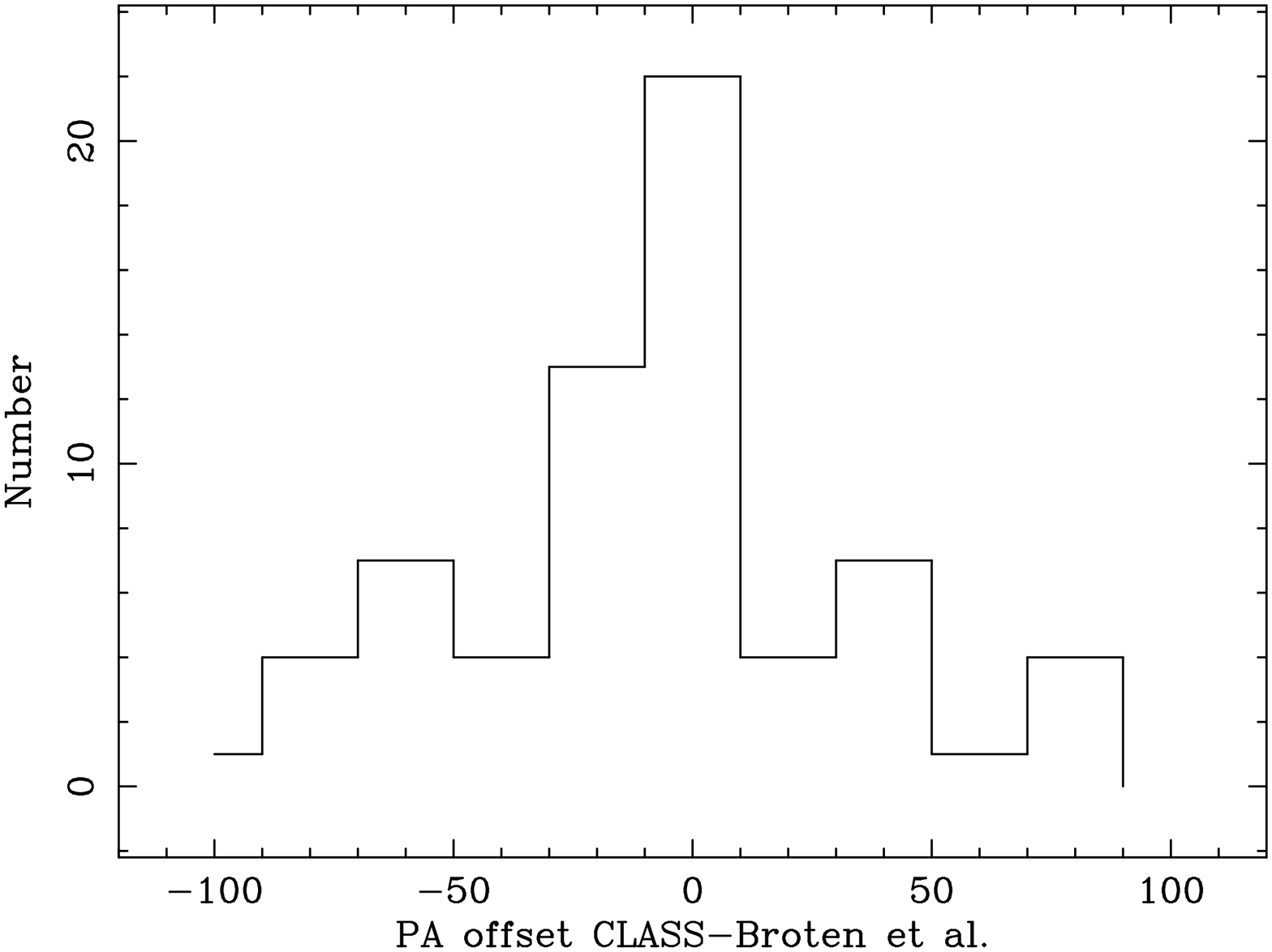,width=8cm,angle=-90}\\
\psfig{figure=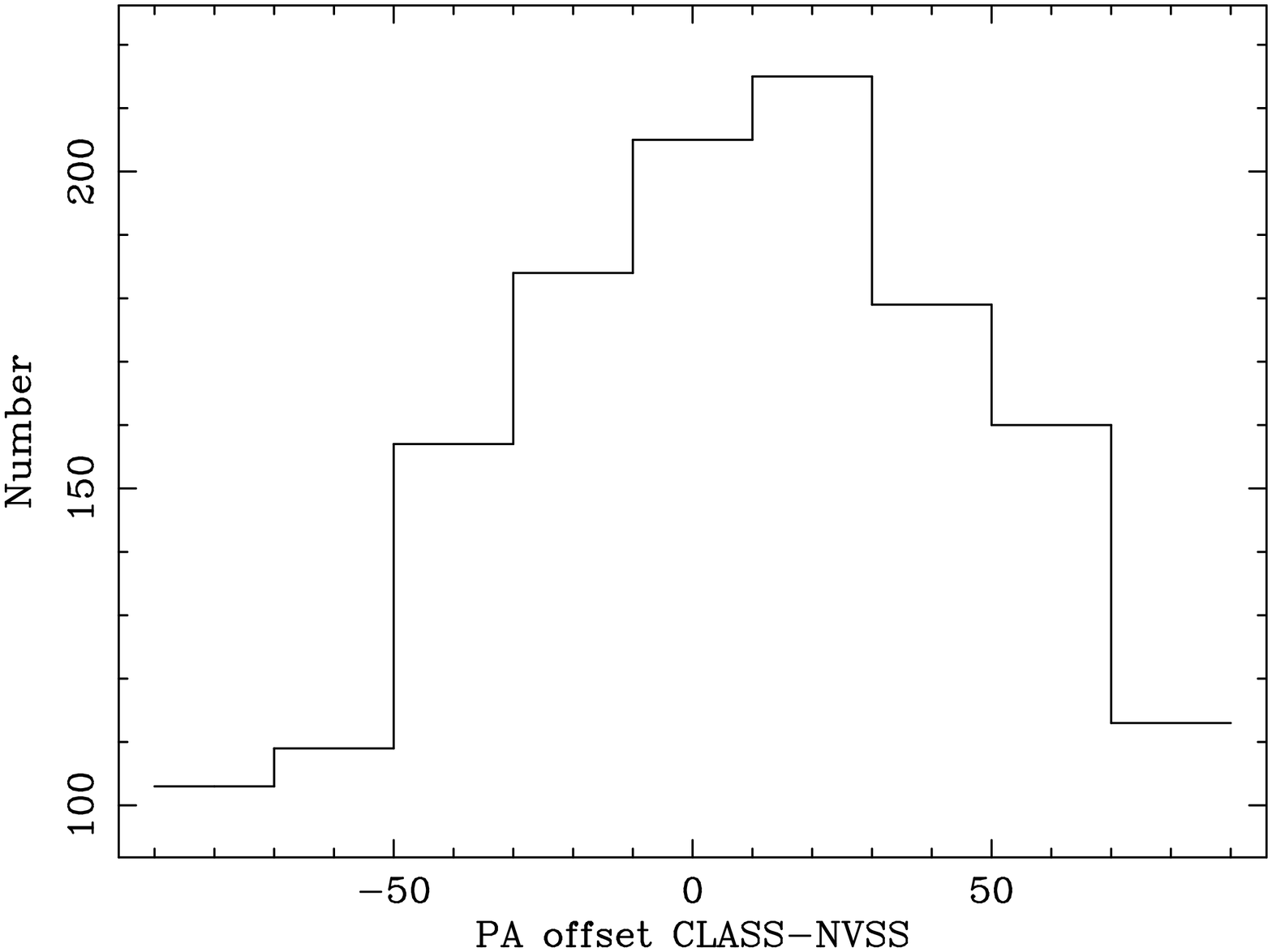,width=8cm,angle=-90}\\
\end{tabular}
\caption{Histogram of position angle differences, in degrees, 
between CLASS and other
surveys. From top to bottom: Okudaira et al. (1993), Broten et al.
(1988) and NVSS (Condon et al. 1998). In the case of NVSS we have
required sources with significant polarization in each survey 
($>$1~mJy for CLASS and $>$2~mJy for NVSS).}
\end{figure}

The offset in the peak from 0$^{\circ}$ in the CLASS-NVSS polarization
angle difference is almost certainly due in part to large-scale 
rotation measure structure in the Galaxy. Simard-Normandin \& Kronberg (1980) present
RM measurements over the whole sky, and identify regions of
60$^{\circ}$--90$^{\circ}$ in extent with correlated Galactic RM. To
demonstrate that the observed offset in the peak is probably due to 
Galactic Faraday rotation, we have selected sources in regions found to have
significant positive RM (defined by $90^{\circ}<l<165^{\circ}$,
$b>0^{\circ}$; Simard-Normandin \& Kronberg 1980) 
and significant negative RM (defined by $\delta>0^{\circ}$,
$l<130^{\circ}$ and $b<0^{\circ}$, corresponding to Simard-Normandin \&
Kronberg's region A). Plotting these separately (Fig. 4) clearly shows
that the polarization differences are offset in different directions in
these two regions of sky.

Finally we emphasize that seeing any peak in the CLASS-NVSS position
angle histogram implies that the average integrated rotation measures of the
sources in the sample must be $\leq$ 20 rad m$^{-2}$. This is somewhat
surprising since in these flat-spectrum sources it is believed that
most of the emission originates on parsec-scale structures, and it is
now well established that the core regions detected in cm-wavelength
VLBI observations typically have rotation measures of several hundreds
of rad$\,$m$^{-2}$ (e.g. Zavala \& Taylor 2004). Indeed, the core rotation
measures revealed by 7mm-2cm VLBI observations sometimes approach
1000~rad$\,$m$^{-2}$ or more (Mutel et al. 2005; Gabuzda et al. 2006).


\begin{figure*}
\begin{tabular}{cc}
\psfig{figure=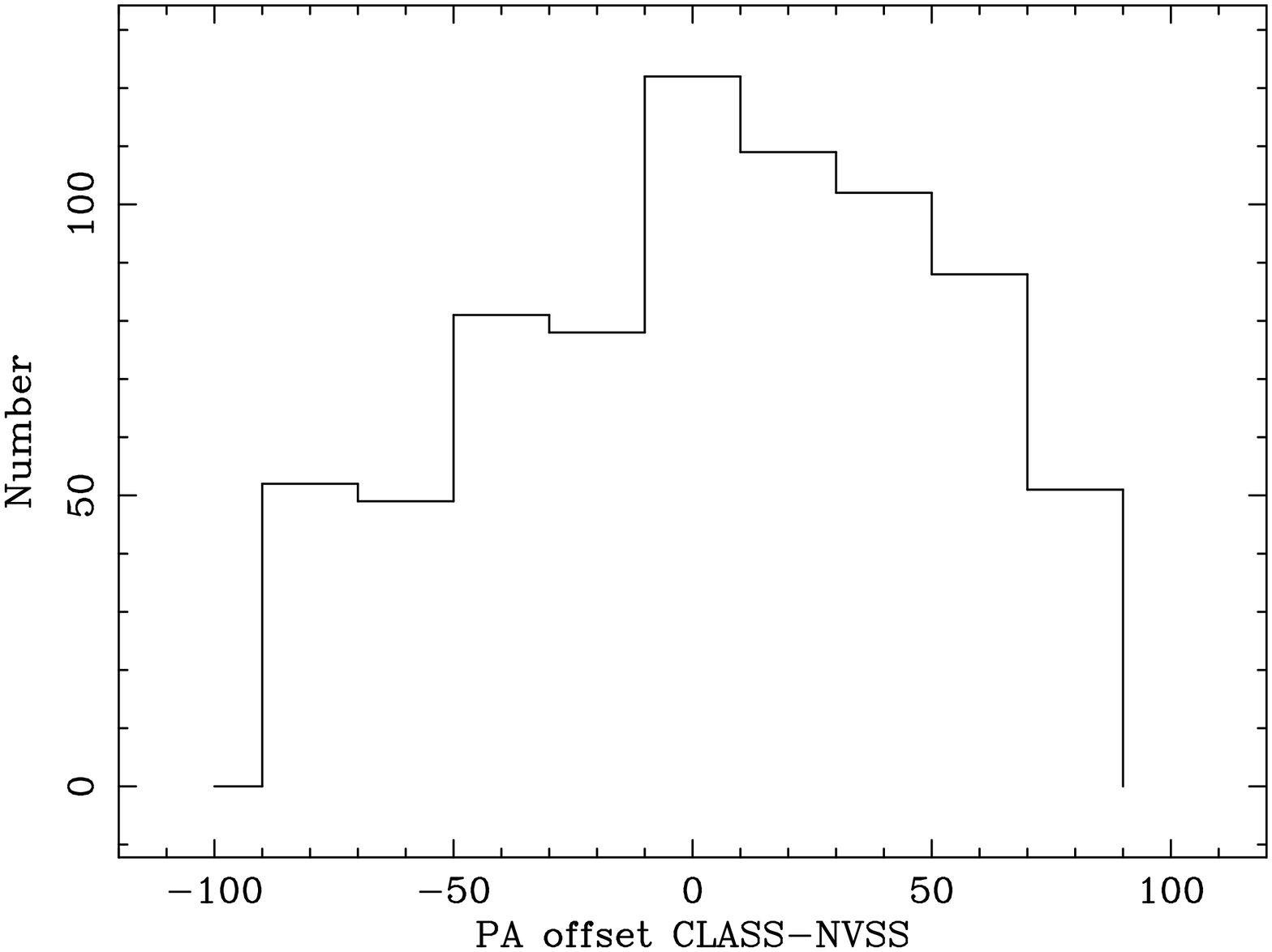,width=8cm,angle=-90}&
\psfig{figure=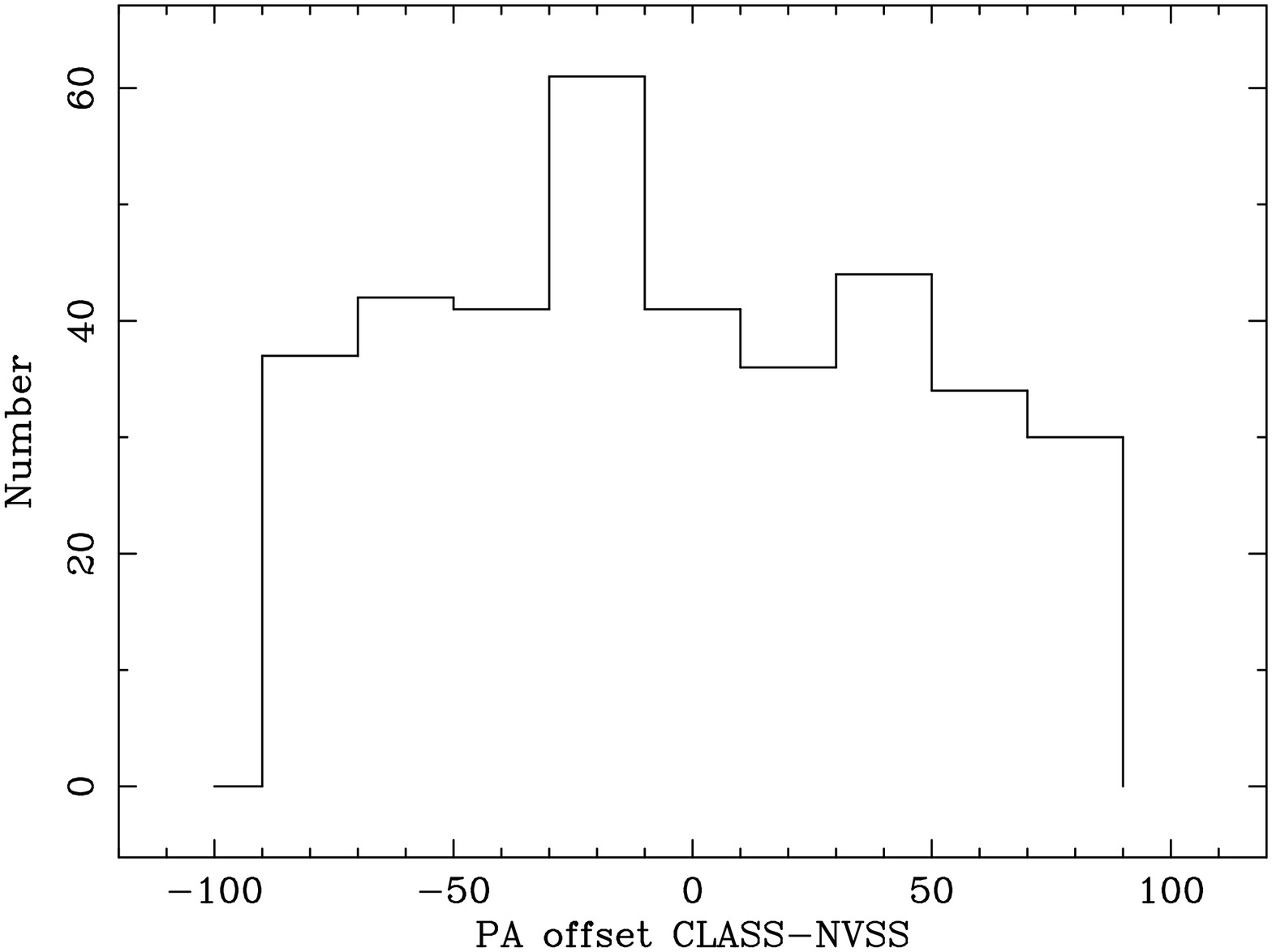,width=8cm,angle=-90}\\
\end{tabular}
\caption{CLASS-NVSS polarization position angle differences for two
regions of sky. On the left is the region $90^{\circ}<l<165^{\circ}$,
$b>0^{\circ}$, found to have significant positive RM by Simard-Normandin
\& Kronberg (1980). On the right is the region of significant negative
RM, $\delta>0^{\circ}$, $l<130^{\circ}$ and $b<0^{\circ}$.}
\end{figure*}

We have also performed an indirect test. It is known both statistically, and
from detailed mapping of individual radio sources, that there is a
correlation between the structure of radio jets and the magnetic field
direction inferred from radio polarization measurements. In particular
high-luminosity quasar jets tend to have dominant longitudinal
magnetic fields (Bridle \& Perley 1984; Cawthorne et al.
1993). Therefore we might expect to see a correlation between jet
position angles and polarization angles. We have compiled a list of 
157 sources which have VLBI elongation directions given in the tables of 
Taylor et al. (1994) and Henstock et al. (1995).
In Figure 5 we show the histogram of the difference between
structural and polarization position angles for these sources in
common between the two samples. There is a clear excess around 90
degrees difference, as expected if magnetic field direction aligns
with that of the radio jet.

\begin{figure}
\psfig{figure=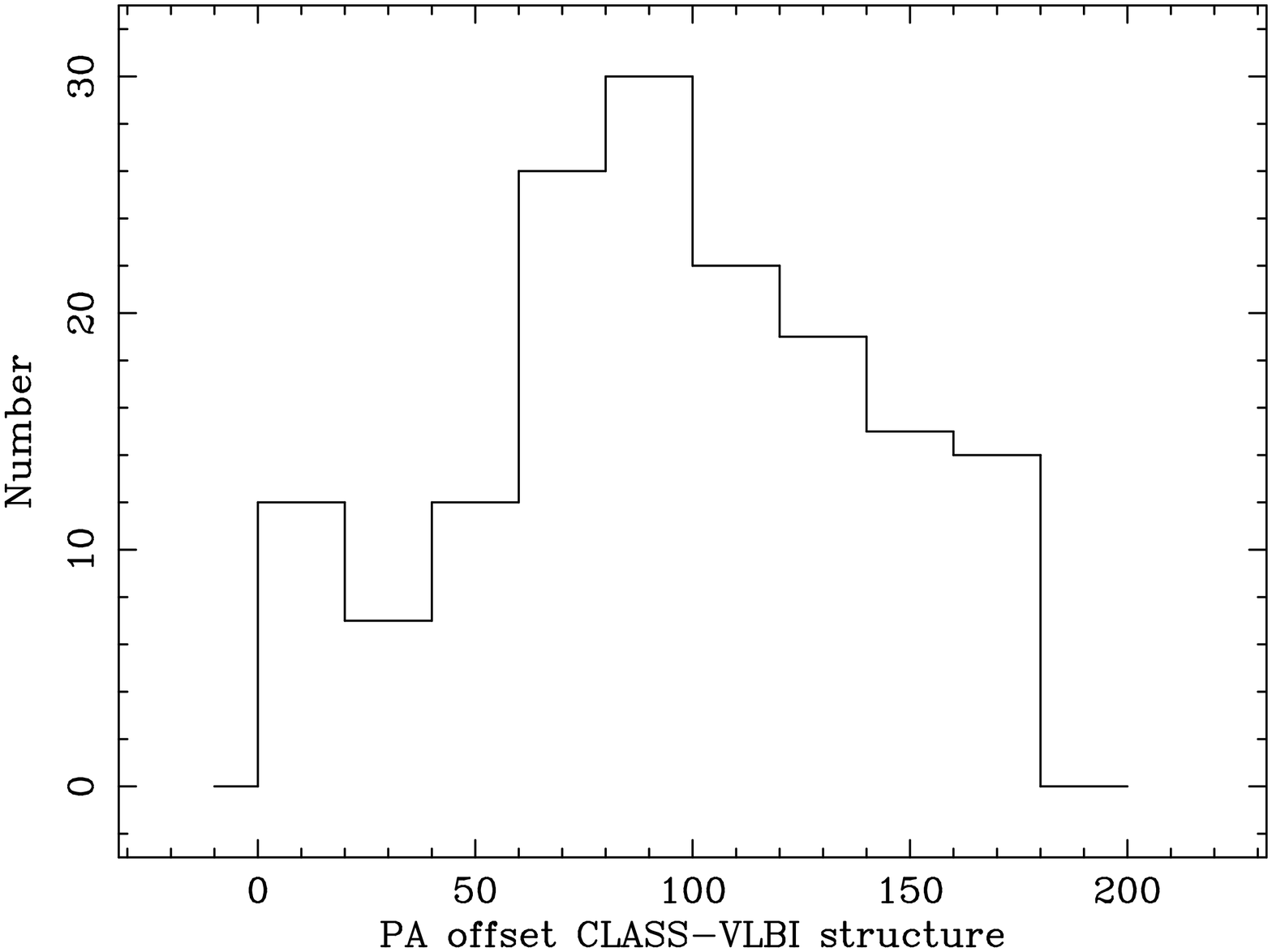,width=8cm,angle=-90}
\caption{Histogram of position angle offsets, in degrees, between CLASS
polarization vectors and VLBI structure from the observations of
Henstock et al. (1995) and Taylor et al. (1994). Note the peak at about
90$^{\circ}$.}
\end{figure}

\section{Conclusions}

We have presented polarization determinations for the 76\% of
CLASS sources  for which reliable polarization calibration is
possible. Although systematic effects are important in determining low
levels of polarization, we have quantified the systematic errors
involved and compared the data with other surveys. Agreement is
obtained which convinces us that the polarization angles are reliable
to within a few degrees for sources with a significant ($>$1~mJy)
polarized flux density.

\section*{Acknowledgements}
We thank numerous colleagues for advice, particularly Paddy Leahy, Ger
de Bruyn and 
Simon Garrington. Jacques Vall\'{e}e kindly provided us with an
electronic version of the Broten et al. polarization catalogue. We thank
the referee, Denise Gabuzda, for useful comments. The
National Radio Astronomy Observatory is a facility of the National
Science Foundation operated under cooperative agreement by Associated
Universities, Inc.  This research has made use of the NASA/IPAC
Extragalactic Database (NED) which is operated by the Jet Propulsion
Laboratory, California Institute of Technology, under contract with
the National Aeronautics and Space Administration.  This work was
supported in part by the European Community's Sixth Framework Marie
Curie Research Training Network Programme, Contract No.
MRTN-CT-2004-505183 "ANGLES". S.J. thanks the University of Manchester
School of Physics and Astronomy for support. Finally, we acknowledge
the vital role played by A.R. Patnaik in the original JVAS survey.

\section*{References}

\noindent Aller M.F., Aller H.D., Hughes P.A., 2003. ApJ 586, 33.


\noindent Birch P., 1982. Nature 298, 451.

\noindent Bridle A.H., Perley R.A., A  1984., ARA\&A 22, 319.

\noindent Broten N.W., MacLeod J.M., Vallee J.P., S  1998.Ap\&SS 141, 303.

\noindent Browne I.W.A., Wilkinson, P.N., Patnaik, A.R. Wrobel, J.M., 1998, MNRAS, 293, 257.

\noindent Browne I.W.A., Wilkinson P.N., Jackson N.J.F., Myers S.T., Fassnacht C.D., Koopmans L.V.E., Marlow D.R., Norbury M., Rusin D., Sykes C.M.,  2003. MNRAS 341, 13.

\noindent Carroll S.M., Field G.B., 1997.  PhRvL 79, 2394.

\noindent Cawthorne T.V., Wardle J.F.C., Roberts D.H., Gabuzda D.C., Brown L.F., 1993. ApJ 416, 496.

\noindent Condon J.J., Cotton W.D., Greisen E.W., Yin Q.F., Perley R.A., Taylor G.B., Broderick J.J., 1998. AJ 115, 1693.

\noindent Ferrari A.A., 1998., ARA\&A 36, 539.

\noindent Fraix-Burnet D., 2002. A\&A 381, 374

\noindent Gabuzda D.C., Rastorgueva E.A., Smith P.S., O'Sullivan S.P., 2006. MNRAS 369, 1596

\noindent Garrington S.T., Leahy J.P., Conway R.G., Laing R.A., 1988. Nature 331, 147.

\noindent Gizani N.A.B., Leahy J.P., 2003. MNRAS 342, 399.

\noindent Gregory P.C., Condon J.J., 1991. ApJS 75, 1011.

\noindent Gregory P.C., Scott W.K., Douglas K., Condon J.J., 1996. ApJS 103, 427

\noindent Hardee P.E., 2003. ApJ 597, 798.

\noindent Henstock D.R., Browne I.W.A., Wilkinson P.N., Taylor G.B., Vermeulen R.C., Pearson T.J., Readhead A.C.S., 1995.  ApJS 100, 1.


\noindent H\"ogbom J.A., 1974. A\&AS 15, 417.

\noindent Hughes P.A., 2005. ApJ 621, 635.

\noindent Hutsem\'ekers, C., Cabanac, R., Lamy, H. Sluse, D., 2005, A\&A, 441, 915.


\noindent Laing R.A., 1988. Nature 331, 149.

\noindent Laing R.A., 1996. Energy transport in radio galaxies and
quasars, Astronomical Society of the Pacific Conference Series, Vol. 
100, Tuscaloosa, Alabama, 19-23 September 1995, publ. Astronomical 
Society of the Pacific (ASP), San Francisco, eds. Hardee P.E. et al., p. 241

\noindent Laing R.A., Canvin J.R., Cotton W.D., Bridle A.H., 2006. MNRAS 368, 48

\noindent Leismann T., Ant\'on L., Aloy M.A., M\"uller E., Mart\'{\i} J.M., Miralles J.A., Ib\'a\~nez J.M., 2005. A\&A 436, 503.

\noindent Lister M.L., Homan D.C., 2005. AJ 130, 1389.

\noindent Lyutikov M., Pariev V.I., Gabuzda D.C., 2005. MNRAS 360, 869.

\noindent Mutel R.L., Denn G.R., Dreier C., 2005. Future directions in
high-resolution astronomy, A.S.P. Conf. Ser. vol 340, eds Romney J.D.,
Reid M.J., p. 155

\noindent Myers S.T., Taylor G., 2006 VLBA Scientific Memo 26, NRAO.

\noindent Myers S.T., Jackson N.J., Browne I.W.A., de Bruyn A.G., Pearson T.J., Readhead A.C.S., Wilkinson P.N., Biggs A.D., Blandford R.D., Fassnacht C.D.,  2003. MNRAS 341, 1.

\noindent Nodland B., Ralston J.P., 1997.  PhRvL 78, 3043.

\noindent Okudaira A., Tabara H., Kato T., Inoue M. 1993. PASJ 45, 153.

\noindent Patnaik A.R., Browne I.W.A., Wilkinson P.N., Wrobel J., 1992. MNRAS 254 655.

\noindent Raffelt G., Stodolsky L., 1988. Phys. Rev. D, 37, 1237

\noindent Rengelink R.B., Tang Y., de Bruyn A.G., Miley G.K., Bremer M.N., R\"ottgering H.J.A., Bremer M.A.R.,
1997, A\&AS 124, 259

\noindent Rudnick L., Jones T.W., 1983  AJ 88, 518.

\noindent Saikia D.J., Kodali P.D., Swarup G., 1985 MNRAS 216, 385

\noindent Saikia D.J., Salter C.J., 1988. ARA\&A 26, 93.

\noindent Shepherd M., in ADASS VI ASP Conf., Ser., vol 125 1997. eds Hunt G., Payne H.E., p.77

\noindent Simard-Normandin M., Kronberg P.P., 1980. ApJ 242, 74

\noindent Simmons, J.F.L., Stewart, B.G. 1985. A\&A, 142, 100

\noindent Taylor G.B., Vermeulen R.C., Pearson T.J., Readhead A.C.S., Henstock D.R., Browne I.W.A., Wilkinson P.N., 1994  ApJS 95, 345.

\noindent White R.L., Becker R.H., 1992. ApJS 79, 331.

\noindent Wilkinson, P.N., Browne, I.W.A. Patnaik, A.R., Wrobel, J.M., Sorathia, B., 1998. MNRAS, 300, 790.

\noindent Zavala, R.T, Taylor, G.B., 2004. ApJ, 612, 749

\bigskip
\end{document}